# On a probabilistic definition of time

Alberto Bicego
(albertobicego@libero.it)

A mechanism is proposed that allows to interpret the temporal evolution of a physical system as a result of the inability of an observer to record its whole state and a simple example is given. It is based on a review of the concepts of information, entropy and order. It is suggested that the temporal evolution and the choice of the "initial state" depend on the way the observer "compresses" information.

## *Introduction.*

Carlo Rovelli has recently returned [1] on an issue repeatedly addressed in the past [2–5], namely on the advisability to introduce the concept of time in quantum gravity. His answer is clear: "The answer I defend is that we must forget the notion of time altogether, and build a quantum theory of gravity where this notion does not appear at all."
The "problem of time" is central for any attempt to combine general relativity with quantum mechanics, as is clear at least since Wheeler-DeWitt's equation was proposed, but the positions on its possible solution are very different [6, 7]. However, even if one does not want to adhere to the position of those who consider time an indispensable ingredient in any physical theory [8], it is obviously necessary that this concept can be found at least as a result of some process of limit – similar to the classical limits of relativity or quantum mechanics – or may emerge in some particular situation. The mechanism proposed by Rovelli to recover the notion of time is a thermodynamic origin [1, 2]; Julian Barbour [9, 10] suggests that it is our brain, through certain structures of the universe that he calls *time capsules*, to give us the illusion of a passing time, just like a movie projector from a set of frames.
In addition, in an attempt to address some unsolved problems, such as the collapse of the wave function, Rovelli reaches a reformulation of quantum mechanics based on information theory [11, 12].
I fully endorse the opinion that time is not something flowing someway outside the universe, but it is the result of an interaction between the universe and our brain. I am also convinced that information theory should play an important role in physics not only, for example, in order to obtain quantum computers, but at a more fundamental level; an opinion, moreover, already expressed even more categorically, amongst others, by J.A. Wheeler [13]. I can't say, of course, whether Rovelli's or Barbour's ideas will actually lead eventually to a consistent and complete theory of quantum gravity and I will not deal with this problem. I'll just show how, in some cases, a notion of time can be derived from the probabilistic nature of information recording. According to Anderson's classification [7], I should say that, rather than the "problem of time mismatch", I'll address the problem of the "arrow of time".

## *Entropy.*

I must begin by saying that I will use the terms entropy and information in Shannon's original meaning [14], although there is still a vigorous debate on their meaning and in particular on the possible relationship between different definitions of entropy in various fields. [15–17].
So, consider an event that can be realized in $N$ different ways $x_i$, each with probability $p_i$. Shannon defines *entropy* of the set $p_1, \ldots, p_N$

$$(1) \quad H = -k \sum_{i=1}^{N} p_i \log_2 p_i$$

where $k$ is arbitrary and I have chosen 2 as base of the logarithm, to have the bit as information unit. If we have no knowledge at all about the event, we shall give to all the results the same probability $p = 1/N$ and therefore

$$(2) \quad H = k \log_2 N$$

Let us now consider a physical system $S$. The system $S$ can be known by an observer $O$ through a series of measures that determine its state. The observer $O$ can consider $S$ as the set of possible outcomes of the measurement process.[1] Since we are interested in some kind of record of the state of $S$, we consider a state of $S$ as specified by a certain number, call it $s$, of bits (the number of possible states of $S$ will therefore be $N = 2^s$).

We shall call *measure* any physical interaction between $S$ and $O$ because of which $S$ will be in a state $x$ and $O$ in a state $X$. Ultimately we can consider a measure as a function $m: S \to O$ from $S$ to $O$.

In principle, if the number of bits available to $O$ to carry out the measure is at least equal to $s$, a measure could be a bijection, so that a state of $S$ is uniquely specified by one of $O$'s, but it is clear that in most cases it will not be so: think of a gas, whose state is specified by the knowledge (at least) of the positions and moments of all particles, while measures allow us to know only the macroscopic parameters.[2] We'll call *micro-states* the states $x$ of $S$ and *macrostates* the states $X$ of $O$ (also $O$ will be the thought of as the set of its states). No ambiguity will arise if we call *macrostate* also the set of microstates corresponding to the same state of $O$, i.e. $m^{-1}(X)$; we can thus say that $x \in X$ and call microstates belonging to the same macrostate *equivalent*. The way in which $O$ takes its measurements allows us therefore to build in a natural way a partition on $S$.

Suppose now that $O$ addresses the problem to predict the outcome of a measure: he will give each state a probability $p(x)$ of showing up. $x$ is therefore a random variable and it is possible to introduce for it an entropy in Shannon's sense:

$$(3) \quad H(x) = -k \sum_S p(x) \log_2 p(x)$$

where the sum is extended to all states of $S$.

If $O$ does not know anything about the system $S$, he will have no reason to assign different probabilities to the various microstates, so in the absence of information about the system $S$,

$$(4) \quad H(x) = k \log_2 N = ks$$

It should be noted (as Shannon says explicitly) that $H$, in this approach, is not in any way a function of the state $x$, it is simply a number describing the amount of information that the observer $O$ has on the system $S$: in the case of minimum information, it is equal, up to a constant, to the number of bits used to describe a state of $S$.

Assuming now that the result of a measurement has been $X$, $O$ will be interested to know in which microstate $S$ is. Again, in the absence of other information he will assign to all microstates corresponding to $X$ the same probability

$$(5) \quad p(x|X) = \begin{cases} 1/w(X) & \text{if } x \in X \\ 0 & \text{otherwise} \end{cases}$$

---

[1] One may wonder if $S$ is actually something more than this, i.e. whether it has an existence independent of $O$' observation ("The notion rejected here is the notion of absolute, or observer-independent, state of a system; equivalently, the notion of observer-independent values of physical quantities." Rovelli in [11]), but we will not deal with this problem.

[2] In fact, the existence of a much larger number of states of the system than is possible to measure, is supposed by us on the assumption that the gas is made up of particles; it is therefore shareable the definition given by Rovelli [11], who regards the state of a system as the maximum amount of information one can get from it.

where I called $p(x|X)$ the probability that the state of S is x conditioned by the fact that the measurement result is X and w, according to custom, the multiplicity of the macrostate X. Therefore
(6)     $H(x|X) = k \log w(X)$

It is (6), not (2) that must be interpreted as Boltzmann's equation. A value of H is therefore associated with each macrostate X, but we can also associate this value with each state $x \in X$, so that we can consider H as a function defined on S, although not bijective. We shall therefore write H(x). This function tells us that the higher the multiplicity of the macrostate X to which x belongs, the lower the information on the system S that O has gained by the measurement: only if w = 1 does O know with certainty the state of S.

The meaning of H is here precisely that of information theory, since S can be regarded as a signal source, O the destination and the measuring instrument the channel; so that O, receiving the message X, will raise the question of what the original content x is.

It should be noted that in this approach *the definition of entropy makes sense only in reference to a specific observer* and has no meaning for the system itself. *Different observers can, in principle, construct different entropy functions in relation to different ways of grouping states together.*

## *Time evolution.*

Identifying the time evolution of a system means, first of all, being able to order its states, i.e., given two states, say which comes first. All the classic examples make use of an idea like this: one imagines a camera that records the state of the system and notes that he can easily tell whether the movie runs forward or backward.

It must therefore be possible to establish a function which associates with each state of the system a value *t* of a real parameter. In fact, the function may not be injective, since all we can measure are the macrostates. It is unavoidable the temptation to nominate entropy itself as a possible "time" of the system; but any increasing function of H will do. That solves the problem of ordering the states of S, but still it is not enough.

It is usally said that a physical system evolves naturally over time so as to increase its entropy as it passes from less probable states to more probable ones. Actually we are referring, of course, to macrostates, the only ones we can really observe.

The idea underlying this explanation of the temporal evolution is that all microstates are equally likely and so it seems natural that the temporal evolution goes in the direction of increasing multiplicity. On closer reflection, however, this view appears to be an undue extension of the fundamental axiom of statistical mechanics. The axiom of microstates equiprobability concerns only states of equilibrium and excludes the possibility of temporal evolution; indeed, statistical mechanics completely ignores the idea of time. "Statistical mechanics, however, does not describe how a system approaches equilibrium, nor does it determine whether a system can ever be found to be in equilibrium. It merely states what the equilibrium situation is, for a given system" (Kerson Huang in [18]).

Besides, it does not justify the observation that evolution is progressive: if we observe that a system at a given time is in a state with low multiplicity (not likely), we can also expect that at a later moment the system state belongs to a macrostate with higher multiplicity, but this hardly explains why the steady state is reached gradually. If all microstates are equally likely, why is the system not in a state of high multiplicity immediately after the first measurement?

It seems to me that a system could present a time evolution only if its states are not equally probable. It's impossible to attribute a time evolution to a system, if, at any time, it has the same probability of being in any microstate; the idea of time evolution is inextricably linked to a gradual change.

So, let's see how we can introduce a *time* in S.

We will need the following ingredients:

- a partition of S into macrostates, as described above, with these features:
  - there is a subset $O'$ of $O$ in which different macrostates have different multiplicity, so that
  
(7) $\quad X_i \neq X_j \rightarrow H(X_i) \neq H(X_j)$

  - there is a state $X_0 \in O'$, which we'll call *initial (macro)state*, having

(8) $\quad H(X_0) = 0$

  and thus multiplicity 1. We'll call *initial (micro)state* also $x_0 = m^{-1}(X_0)$;

- a bijection $t: O \rightarrow I \subset R^+$ which is an increasing function of $H$;
  from (6) follows that $t$ takes its minimum value in $X_0$: we'll set

(9) $\quad t(X_0) = 0$

- a function $p: O \times I \rightarrow [0;1]$ with these features:
  - for each value of $t$, $p$ is a probability function defined on $O$; in particular we'll have therefore:

(10) $\quad \sum_O p(X,t) = 1 \quad$ for each fixed $t$ (the sum is over all the macrostates of $O$);

  - initially the system is certainly in $x_0$:

(11) $\quad p(X_0, 0) = 1$

  - for each fixed $X$ the function $p(X,t)$ has one maximum in $t_m = t(X)$:

(12) $\quad \max(p(\overline{X},t)) = t_m \leftrightarrow t_m = t(\overline{X})$

In this case *we'll say that t is a time for the observer O in relation to the set of states O'*.

It may seem strange that $t$ is a function defined on $O$ and not on $S$, but remember that the states of $S$ are inaccessible to $O$: all he can know are the macrostates $X$.
$t$ can anyway be traced back to be a function on $S$ as follows:

(13) $\quad t(x) = t(m(x))$

Of course $t(x)$ is not an injective function.
Also $p$ can be made to be a function on $S \times I$, defining

(14) $\quad p(x,t) = \frac{1}{w(X)} p(X,t) \quad$ with $X = m(x)$

In this way the probabilities of two microstates belonging to the same macrostate are the same and each one reaches the maximum at the same $t$ of the corresponding $p(X,t)$.

We can therefore speak of a time for S, but we must not forget that this makes sense only in reference to an observer $O$: *an observer that constructs different macrostates might observe a different time evolution*.

## *An example.*

It may seem that the proposed mechanism is artificial and complicated, but we'll see an example in which the above procedure reveals simple and effective.
Consider a system S whose state $x$ is specified by a number $s$ of bits. We can imagine a state of the system as a series of boxes s filled with 0 and 1. Suppose we are not able to record a full state $x$, but only the number $n$ of zeros it contains. The number $n$ specifies then the macrostates of the system. The multiplicity of $n$ is

(15) $\quad w(n) = \frac{s!}{n!(s-n)!}$

and the corresponding entropy

(16) $\quad H(n) = k \log_2 \dfrac{s!}{n!(s-n)!}$

$O$ is, for us, the set of values of $n$. Observe that there are different values of $n$ with the same multiplicity, e.g. $n = 0$ and $n = s$. But if we consider $O' = \{n \,|\, n < s/2\}$, then all the macrostates have different multiplicity and $H$ is an increasing function of $n$.

Let's introduce now the function

(17) $\quad t = \log_b \left(1 - 2\dfrac{n}{s}\right) \qquad$ where $\quad b = 1 - 2p$

and $p$ is an arbitrary parameter between 0 and 1/2 (excluding extremes). Let's introduce, besides, the function[3]

(18) $\quad p(n,0) = \delta_{n0} \qquad\qquad\qquad\qquad\qquad\qquad\qquad$ if $\quad t = 0$

$\quad p(n,t) = w(n) \dfrac{\left(1 + (1-2p)^t\right)^{s-n} \left(1 - (1-2p)^t\right)^n}{2^s} \qquad$ if $\quad t > 0$

It is easily verified (see Appendix 1) that (7–12) are satisfied.

Note that $S$ does not have, per se, any intrinsic notion of time, but the way in which $O$ stores the states of $S$, i.e. the way in which he makes a measurement, authorizes us to consider the function $t$ a time for $O$. He will observe that by making a measurement on $S$ he will have a different probability of obtaining the various states while the "time" $t$ is changing. As an example, I show in *Figure 1* some values of $p(n,t)$. The system, therefore, according to $O$, "evolves" from the initial state in which all the bits have value 1 to a final state in which half of the bits (as a limit value, since for $n = 10$ the curve has no maximum) are 0.

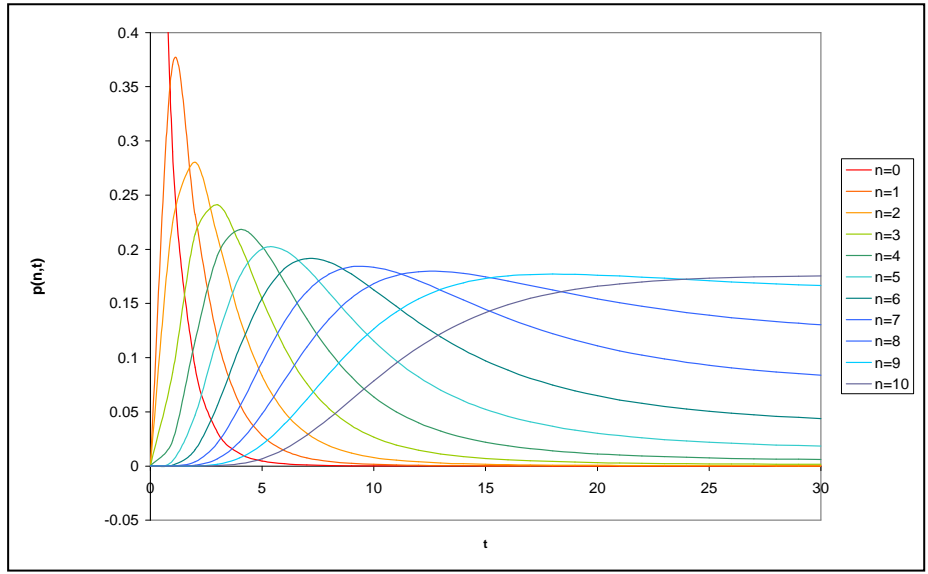

**Figure 1**: The probability for $O$ of finding $n$ zeros depends on the "time". The system thus "evolves" from an initial state where all the bits are 1 to a final state in which half of the bits are 0. In this graph I have set $p = 0.06$ and $s = 20$, with $s$ number of zeros and $p$ is a scaling factor for $t$.

The "*trajectory*" of $S$, as seen by $O$, can be obtained simply by inverting (17):

**(19)** $\quad n(t) = \dfrac{s(1-b^t)}{2}$

---

[3] One might wonder where these functions come from. I derived them for an earlier work that was not published. This does not matter at all here: we shall take (17) and (18) as simple definitions: their adequacy will be justified by the results we obtain.

The parameter *p* plays the role of scale factor for *t*, and determines, therefore, its unit.

One might think that this is an entirely abstract contruction and that *S* does not evolve at all, however, this model describes just one of the examples that are used to explain the meaning of entropy and time evolution: the urn model, proposed in 1907 by P. and T. Ehrenfest [19]. This is a box divided into two parts; initially only one of them contains a number *s* of particles; the box is shaken and every second a particle has a probability *p* of jumping on the opposite side. Assigning to the compartment which initially contains the particles the value 1 and to the other 0, the system is described by *s* bits that specify the position of each particle. The time evolution is precisely that described above, which we derived considering solely how *O* stores the information about *S*.

## *Order and reversibility.*

Let's try to understand better the relationship between information and temporal evolution. It is usually said that a system evolves spontaneously from order to disorder. States of low entropy are considered *ordered* and states of high entropy *disordered*. You can take the example of a deck of playing cards, originally prepared "in order" and then mixed repeatedly. Everyone knows that, while in theory possible, in practice there is no hope of finding the initial state again. We need an external intervention to prepare the system in this state, after which the system, if we let it "evolve" spontaneously, will roll away from it relentlessly. Similar considerations were applied to the universe as a whole, inducing cosmologists to ask: "What forced the entropy of our world to be so low in the past? (...) To produce a universe similar to that in which we live, the Creator would have to aim at an absurdly small volume of phase space of possible universes" (Roger Penrose in [20]).

However, returning to our example of the deck, we can ask why the initial state should be so special. Suppose that from an initial state, which we call $x_0$, completely "ordered", shuffling a large number of times we have come to a state fully "disordered"; in what are $x_0$ and $x_1$ different? All in all, if we mix, we will not expect to find $x_1$ more than to find $x_0$. Imagine turning the cards, which are in the state $x_1$, face down and write on each one a progressive number; they appear now completely "ordered", but if we restore $x_0$ they will appear completely "disordered". Why are we not willing to consider $x_1$ as the initial state without writing the numbers on the back? The answer is that mixing from $x_1$ we would be unable to distinguish any evolution, because the states *are all alike*, or, more simply, because it is not easy to memorize $x_1$.

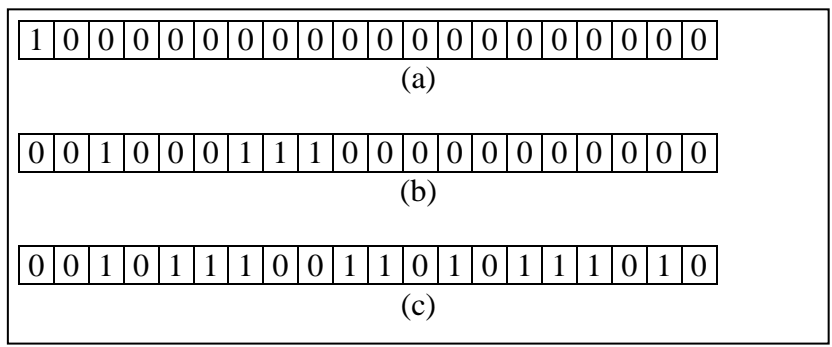

**Figure 2**: A configuration in which few details emerge from a background, such as (a) or (b) is easier to memorize than one in which they become numerous, probably because in the process of recording we use a sort of "compression algorithm", rather that store all the individual bits.

As another example, consider *Figure 2*: it shows three different configurations of a hypothetical system specified by 20 bits. Try to look at each one for a few seconds and then try to recall them. It is quite obvious that while the first (a) is recorded immediately, the second (b) requires greater attention, while the third (c) is more difficult to remember. Generally we find it easier to recognize configurations in which few elements emerge from a homogeneous background, while, if their number grows too large, they tend to merge and become themselves an indistinguishable background.

The reason is probably that in the process of storing our brain takes some strategy of "compression" to reduce the number of bits needed to reproduce the figure. We can, for example, imagine a strategy of this kind: we record the number of consecutive 0, starting from the first box, then the number of consecutive 1, then the number of 0 ... and so on, until the end of the boxes. You can see (Appendix 2) that this method is efficient when the 0s are "compact", but it requires an exaggeratedly high number of bits when they are distributed evenly. It is then possible that our brain gives up a faithful storage, losing information on the system. We saw in the previous example that incomplete recording of the state of a system can lead to attribute temporal evolution.

Now imagine an object on a segment of length $L$; let $l << L$ be the smallest interval that we can appreciate, so that we can imagine our segment as a set of boxes that are occupied or empty; to fix ideas we will say that each box can be white or black. Suppose that our object is long $nl$ ($l << nl << L$) and fills $n$ boxes. Then a state of our system can be described by $s = L/l$ bits - putting, for example, 0 where the box is black - $n$ of which are consecutive zeros. Of course we can use the compression algorithm described above, saving a large amount of memory. The representation of data, in this case is *lossless*, entropy cannot be defined, nor a time evolution for the system in the way described above. But if our object can be separated into particles of length $l$, the restriction that the zeros are consecutive disappears and data compression quickly becomes inefficient for configurations where the particles are finely distributed. It may be convenient then to give up a faithful memorization and use a probabilistic representation. For example, rather then referring to individual white or black boxes, we could build groups and consider, instead of the actual configuration of each group, the "degree of gray", specified by the number of black boxes in the group. It is clear that the reconstruction of the system state will be approximate and we can only "guess" what it really is. In this case it should be possible, somehow, to introduce a time function for the system.

This would suggest that in those which are referred to as *irreversible* phenomena there is a loss of information due to failure to record completely the state of the system, while the ability to know fully the state of the system leads to *reversible* phenomena, for which it is not possible to identify any temporal order.

## *Final remarks.*

The observer $O$ in our example, just because he stores information about $S$ in a certain way, is led to introduce an order in the set of the states of $S$. He can associate with each state a number $t$ and, choosing to use $t$ as a time for the system, he will find that by making a measurement at a given time $t$, he will have a maximum probability of finding some (macro)state. It is easy to see, with a numerical simulation, that the maximum of *Figure 1* become very sharp increasing the number of bits $s$, so that the temporal evolution of the system is well defined.

This mechanism poses some problems, a few of which I will try to highlight.

- Among the ingredients to build a time evolution I put a function $t: O \to I \subset R^+$. It may seem that this means introducing a parameter from the outside, as in Newtonian physics and quantum mechanics. Actually, the central role is played by the function $p(X,t)$: it is maximizing it that you can identify the function $t$. But how can we find $p(X,t)$? In the example it was found, in fact, so as to reproduce the correct time evolution of the system under consideration. One may ask whether there is a general way to determine it. The problem seems somewhat similar to that of finding the action or Hamilton's characteristic function: it is fairly easy, in some cases where one already knows the dynamics, to find these functions, but it is not possible to find them in general, in order to derive the dynamic as a result.

- I suggested that the temporal evolution, as a result of the grouping of states of *S* into macrostates, is due to the inability of our brain to record all the information contained in *S*. Can we hazard a psychological mechanism by which the feeling of time evolution takes place?
Suppose we have a hundred strings like those shown in *Figure 2*, which we are required to reproduce after observing them for some time and that we get a reward for each configuration correctly reproduced. Where shall we start from? No doubt from those, such as *a* and *b*, which we are able to memorize easily and with certainty, while we shall be induced to leave last those for which we can only make attempts. Can we accept that there be some sort of advantage in *paying greater attention* to unmistakable configurations and that we can interpret this as *considering preceding*? In this case Barbour would be right in saying that "it is our brain that plays the movie" [10], giving us the impression of passing time.

- I insisted that the time evolution of a system depends on how an observer records the system. Consider a configuration in which the bits have alternate values: 010101010 ... Of course if we wanted to use the compression algorithm described above it would be a bad deal, but everybody will agree that this can be considered a "highly ordered" state, since it is very easy to memorize. It can easily be traced back to the case where the first half of the bit is 0, simply changing the order in which one counts the boxes. If we associate, as before, a black box to 0 and white one to 1, we have a something that can roughly represent the diffusion of two liquids of different colors; it follows that if it is possible to identify for it a time evolution, it should also be possible to have a "reversed" evolution, starting from a situation in which the molecules of the two liquids are evenly distributed. But we must believe that this way of counting is not spontaneous, or it is not convenient for our brain.

## *Appendix 1.*

I'll show that the example proposed satisfies the conditions (7–12).
Since the initial macrostate is $n = 0$, (7–9) are immediate.

(10) becomes $\sum_{n=0}^{s} w(n) \frac{(1+b^t)^{s-n}(1-b^t)^n}{2^s} = \frac{1}{2^s} \sum_{n=0}^{s} \binom{s}{n}(1+b^t)^{s-n}(1-b^t)^n = 1$ for the binomial formula.

To prove (12) it suffices to annul the derivative respect to $t$ of $(1+b^t)^{s-n}(1-b^t)^n$ (or its logarithm) and note that (17) follows. $H$ is an increasing function of $t$ because $w$ is an increasing function of $n$ (for $n < s/2$) and $n$ is an increasing function of $t$, as can be seen from (19) (remember that $b < 1$); its expression follows from (16) e (19):

$$H(t) = k \log_2 \frac{s!}{\left(\frac{s(1-b^t)}{2}\right)! \left(\frac{s(1+b^t)}{2}\right)!}$$

## *Appendix 2.*

We want to memorize the state of a system of $s$ bit.
Consider the following "compression algorithm":

- read the first bit: if it is 0 write 0, if it is è 1, write 1; this requires 1 bit of memory
- count the $n_0$ consecutive bits equal to the first, write $n_0$; since $1 \leq n_0 \leq s$ (including the first bit), this requires about $\log_2 s$ bits
- count the next $n_1$ consecutive bits different from the first one, write $n_1$; since $0 \leq n_1 \leq s - n_0$ this requires about $\log_2(s - n_0)$ bits
- count the next $n_2$ consecutive bits equal to the first one, write $n_2$; since $0 \leq n_2 \leq s - n_0 - n_1$ this requires about $\log_2(s - n_0 - n_1)$ bits
- proceed until you finish the $s$ bits

We have used in this way a number of bits

(A2.1)    $n = 1 + \sum_{k=0}^{c} \log_2 \sum_{i=k}^{c} n_i$

where $c$ is the number of changes in the value of two contiguous bits and $\sum_{i=0}^{c} n_i = s$.

In the sum (A2.1) the most important terms are those for which $\sum_{i=k}^{c} n_i \approx s$, therefore roughly

(A2.2)    $n \approx 1 + a \log_2 s$

where $a$ is a number close to the number of groups of bits that are small.
It is clear that this algorithm works well if the values are grouped into large homogeneous groups, otherwise it fails. We can find a limit setting in (A2.2) $n \leq s$ and solving for $a$; ignoring 1 compared to $s$, we find $a \leq \frac{s}{\log_2 s}$.

We can therefore say that when the bit value changes more than $\frac{s}{\log_2 s}$ times, the compression method becomes useless, because we need more than *s* bits.

Of course I do not want to maintain that the method described is that used by our brain to store information, but it is likely that it implements a compression algorithm of some kind and that when the space required becomes excessive, it prefers, as they say in computer science, a *lossy* procedure, reaching only a probabilistic knowledge of the system it is observing.

## *Appendix 3.*

The system *S* considered in the example has no other feature than to be described by *s* bits. It could therefore represent any physical system. We can amuse ourselves by deriving a "cosmological" result, imagining that *S* represents the entire universe. We will base ourselves on two assumptions (not so easy to accept, indeed):
- the multiplicity of the states represents somehow the "volume" of the universe;
- our perception of time is somehow analogous to that which comes from knowledge of the "number of zeros" of the universe, so that we can use (15−19).

From (15, 16) we have

(A3.1) $$w = 2^{H/k}$$

We can obtain an approximate expression of *H* using Stirling' formula, $\ln x! \approx x \ln x - x$, that holds when *n* and $s - n$ are large. We get:

(A3.2) $$H \approx k \log_2 \frac{s^s}{n^n (s-n)^{(s-n)}}$$

and therefore

(A3.3) $$w(t) \approx \frac{s^s}{n^n (s-n)^{(s-n)}}$$

where *n* is given by (19).

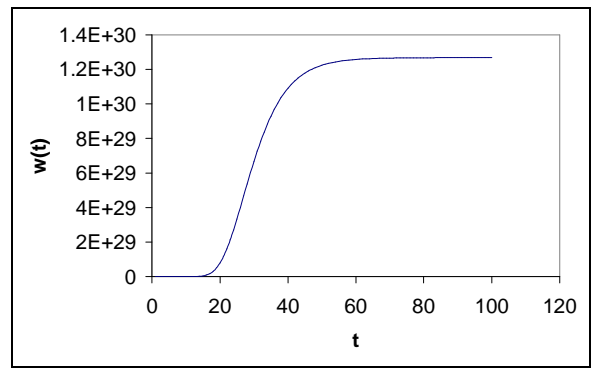

**Figure 3:** Universe's "volume" versus time (in arbitrary units). The universe undergoes a growth of many orders of magnitude, which brings it rapidly close to the maximum size, then slows down to become asymptotically stationary. Here the number *s* of bits that describe the universe is set equal to 100 and the parameter *p*, which determines the time scale, equal to 0.035; *s* determines the asymptotic value of the volume.

*Figure 3* is a graph of *w(t)*. Curiously, in this model the universe would undergo a period of inflation, during which it grows many orders of magnitude, reaching almost the maximum volume, and then slow its expansion becoming asymptotically stationary. The instant when deceleration begins can be found annulling the second derivative of (A3.3), which is

(A3.4) $$\dddot{w} = w \ln^2 b \left(\frac{s}{2} - n\right) \left[\ln \frac{n}{s-n} - \left(\frac{s}{2} - n\right) \frac{s}{n(s-n)} + \left(\frac{s}{2} - n\right) \ln^2 \frac{n}{s-n}\right]$$

where *w* is given by (A3.3) and *n* by (19). Annulling the term in parentheses one gets an equation that cannot be solved exactly in R.

Though with this mechanism inflation arises naturally, probably it cannot be considered satisfactory, because, as Liddle says in [21], "it must come to an end early enough that the big bang successes are not threatened".

---